\documentclass[a4paper,11pt]{article}
\pdfoutput=1 

\usepackage{jheppub} 

\usepackage[T1]{fontenc} 
\usepackage{physics}

\usepackage{subcaption}
\usepackage{caption}

\title{\boldmath Information propagation in a non-local model with emergent locality}

\author[f]{Kaixin Ji,}
\author[a,b,c,d,e,f,g]{Ling-Yan Hung}

\affiliation[a]{State Key Laboratory of Surface Physics, Fudan University, 200433 Shanghai, China}
\affiliation[b]{Shanghai Qi Zhi Institute, 41st Floor, AI Tower, No. 701 Yunjin Road, Xuhui District, Shanghai, 200232, China}
\affiliation[c]{Department of Physics and Center for Field Theory and Particle Physics, Fudan University, Shanghai 200433, China}
\affiliation[d]{State Key Laboratory of Surface Physics, Fudan University, 200433 Shanghai, China}
\affiliation[e]{Shanghai Qi Zhi Institute, 41st Floor, AI Tower, No. 701 Yunjin Road, Xuhui District, Shanghai, 200232, China}
\affiliation[f]{Department of Physics and Center for Field Theory and Particle Physics, Fudan University, Shanghai 200433, China}
\affiliation[g]{Institute for Nanoelectronic devices and Quantum computing, Fudan University, 200433 Shanghai, China}

\emailAdd{kxji21@m.fudan.edu.cn}
\emailAdd{elektron.janethung@gmail.com}

\abstract{
In this paper, we revisit a ''relatively  local'' model proposed in \cite{Lee:2018hsd}, where locality and dimensionality of space only emerges from the entanglement structure of the state the system is in. Various quantities such as butterfly velocity/ entanglement speed can be defined similarly, at least in the regime where locality is well defined and a light cone structure emerges in the correlation between sites. We find that the relations observed between them in local models \cite{brian} are not respected. In particular, we conjecture that the hierarchy of the interaction over different distances provides different “layers” of light
cones. When long range interactions are sufficiently suppressed, the effective light cones are dominated by linear behaviour with little remnant of non-locality.  This could potentially be used as a physical smoking gun for emergent locality in non-local models. }

\begin{document} 
\maketitle
\flushbottom

\section{Introduction}
    
 What does it mean to live in a "local" space-time? This is by no means a trivial question, particularly in the quantum world. 
There are a lot of important insights accumulated in the literature, based on patterns of entanglement, and their dynamical evolution. 
For example, it is by now a classic observation that the entanglement entropy of ground states
of local Hamiltonians follows the {\it area law}. In the limit that the region concerned is large (but small compared to the total system size), the leading term of the entanglement entropy $S_A$ in the large size limit is proportional to the {\it area} of the boundary surface of the given region $A$\cite{Ryu:2006ef}.
 Moreover, in a generic state out of equilibrium (i.e. not an eigenstate of the Hamiltonian), the unitary evolution under the effect of a local Hamiltonian ensures that information can only propagate locally bounded by a finite speed. This is often called the information speed $v_I$. A very similar quantity is the entanglement speed $v_E$.  It is observed that entanglement propagates like a tsunami and the {\it wavefront} moves at a finite speed \cite{tsunami}.  Not surprisingly, in a generic local theory these quantities have simple relations \cite{brian}.  These speed limits connect locality with causality of space-time. 
Chaotic behvaiour is also constrained by locality and causality. A measurement of chaotic behaviour is the famous out-of-time-ordered correlator (OTOC), which essentially is a measurement of the growth of commutator, and thus the analogue of the divergence of {\it nearby paths} under Hamiltonian evolution (See for example \cite{Maldacena:2015waa} and references therein). The growth of the OTOC is known to be controlled by the so called Lyapunov exponent at time scales much larger than thermalization time but much smaller than the scrambling time. 
When the evolution is local however, the Lyapunov behaviour would be restricted to a light-cone, beyond which the OTOC does not grow. The growth of the light-cone is governed by the so called butterfly speed $v_B$. It is also observed to be related to $v_I$ and $v_E$ in local theories \cite{brian}. 

These studies therefore provide a lot of intuitions and quantitative guide to the behaviour of local theories. However, when we move beyond the condensed matter setting and attempt to apply these intuitions in understanding the fundamental nature of space-time that we live in, it is known to be inadequate. While we do not yet have a quantum theory of gravity, that addresses the quantum  nature of gravity, there is evidence that points to the non-local nature of quantum gravity if the theory can in fact be formulated consistently. This follows from requirements of covariance so that the Hamiltonian governing gravitational evolution should be non-local. There are many works that discuss the non-locality of gravity and how that it reconciles with our sense of locality, and how quantum information could perhaps be localized in some sense (There are numerous papers on the subject. For some recent discussions, see \cite{Donnelly:2017jcd, Giddings:2018umg, Donnelly:2018nbv,Giddings:2019hjc,Raju:2021lwh}).
While the Hamiltonian itself maybe non-local, our experience of space-time does not involve the full quantum Hilbert space of quantum gravity. In fact, we are almost exclusively experiencing the situation where there are only minute deviations or fluctuations from a classical background that solves the classical Einstein equation, such as the flat background.   
Therefore, perhaps the sensation of locality is emergent, depending on the small subset of states in the full quantum Hamiltonian that we are actually probing.  This possibility was
explored in explicit toy models in \cite{Lee:2018hsd}. 

In the toy models constructed in \cite{Lee:2018hsd} which we are going to review in more detail in the next section, there is a Hamiltonian that is explicitly non-local. However, the states that were studied are very specially chosen so that an emergent notion of dimensionality and locality emerges. Two degrees of freedom are in close proximity if the initial state carries more entanglement between them, but far when there are very little entanglement in between. This is in line with the physical intuition developed in the past decade, that space-time is a manifestation of entanglement \cite{VanRaamsdonk:2009ar,Maldacena:2013xja}. 

In this paper, we would like to inspect the model in greater detail. Specifically, we would like to see if the notion of information speed, entanglement speed and butterfly speed remain well defined, and whether the connection between them observed in a truly local system remains intact where locality is only emergent and state dependent. 

As we are going to see, indeed these notions remain well defined, although remnants of non-locality can be detected.

\section{Relative locality in a non-local model -- a review}
  \subsection{Model}
  This section is a brief review of the non-local quantum model proposed by Sungsik Lee in \cite{Lee:2018hsd,lee2018emergent}. Define first a set of field operators, $\hat{\phi}_i^a$, with  lower indices $i=1,2,..,L$ labeling a set of sites our system lives on, and upper indices $a=1,2,..,N$ labeling different field components at each site. $\pi_i^a$ are their conjugate momenta with commutators $[\phi_i^a,\pi_j^b]=i\delta_{ij}\delta_{ab}$. The basis states $\ket{\phi}$ are simultaneously eigenstates of all field operators with real eigenvalues $-\infty<\phi_i^a<\infty$,
  
  \begin{equation}
      \hat{\phi}_i^a \ket{\phi} =\phi^a_i \ket{\phi}.
  \end{equation}
  They span the full Hilbert space $\mathcal{W}$. The inner products are normalized as delta functions,
  
  \begin{equation}
      \left\langle\phi^{\prime} | \phi\right\rangle=\prod_{i, a} \delta\left(\phi_{i}^{\prime a}-\phi_{i}^{a}\right).
  \end{equation}

We consider an $O(N)$ symmetric subspace $\mathcal{V}$ of the full Hilbert space $\mathcal{W}$, which is spanned by basis states $|T\rangle$.
\begin{equation}
      |T\rangle=\int D \phi e^{-N \sum_{i,j} T_{ij} O_{ij}}|\phi\rangle,
  \end{equation}
They are generated by a  complete  set of $O(N)$ symmetric field products,

\begin{equation}
      O_{ij} \equiv \frac{1}{N} \sum_{a} \hat{\phi}_i^a \hat{\phi}_j^a.
  \end{equation}

  Each basis state is determined by their collective variables $T_{ij}$. To extract these hopping amplitude $T_{ij}$ from states, we define operator
  \begin{equation}
      \hat{\Gamma}_{ij}=\frac{1}{N}\sum_a \hat{\pi}^a_i\hat{\pi}^a_j.
  \end{equation}
  
   Such operators can act on $|T\rangle$ to produce collective variables (Henceforth repeated indices are summed over unless specified),
  
  \begin{equation}
  \label{coupling}
      \hat{\Gamma}_{i j}|T\rangle=\int D \phi\left[2 T_{i j}-4 T_{i k} T_{l j} \frac{\sum_{a} \phi_{k}^{a} \phi_{l}^{a}}{N}\right] e^{-N T_{i j} O_{i j}}|\phi\rangle.
  \end{equation}
  
A general states in $\mathcal{V}$ reads 
\begin{equation}
    |\Psi\rangle=\int DT \Psi(T) |T\rangle
\end{equation}

where $DT=\prod_{i\leq j}dT_{ij}$ are defined along the imaginary axes.
 In this paper we only consider semi-classical states with wavefunction

\begin{equation} \label{eq:semi_class_state}
\Psi(T)= e^{N \bar{P}_{ij}\left(T_{ij}-\bar{T}_{ij}\right)+\frac{\left(T_{ij}-\bar{T}_{ij}\right)^{2}}{2 \Delta^{2}}}.
\end{equation}

This state is semi-classical because in the limit that 
\begin{equation} \label{eq:ineq}
 N \Delta \gg 1,\qquad  \Delta \ll 1,
\end{equation}
the fluctuations of both $T$ and $O$ are suppressed. i.e.
\begin{equation}
    T_{ij} =  \bar{T}_{ij} + \mathcal{O}(\Delta), \qquad O_{ij} = \bar{P}_{ij} + \mathcal{O}(1/(N \Delta)).
\end{equation}
These can be readily observed by inspecting the Gaussian integrals in (\ref{eq:semi_class_state}). The inequalities (\ref{eq:ineq}) have also been emphasized in \cite{Lee:2018hsd}. In other words, to leading order in the double expansion of $\Delta$ and $1/(N \Delta)$, we can freely replace $T$ by $\bar{T}$ and $O$ by $\bar{P}$.

 Lee went on to introduce an $O(N)$ symmetric Hamiltonian 
  
  \begin{equation}
\label{originalHal}
\hat{H}=-R \sum_{i, j} \sum_{a} \left(\hat{\phi}_{i}^{a} \hat{\phi}_{j}^{a}\right)\hat{\Gamma}_{i j}+U \sum_{i} \sum_{a} \hat{\pi}_{i}^{a} \hat{\pi}_{i}^{a}+\frac{\lambda}{N} \sum_{i} \sum_{a, b}\left(\hat{\phi}_{i}^{a} \hat{\phi}_{i}^{a}\right)\left(\hat{\phi}_{i}^{b} \hat{\phi}_{i}^{b}\right).\end{equation} 
  
  The first term is a universal interaction between any two sites with strength given by $\hat{\Gamma}_{ij}$. It is state-dependent and aimed to mimic gravity that the coupling strength is related to the site-to-site correlations, namely the collective variables. The second and third are kinetic terms and self-energy. 
  
   An infinitesimal time evolution on the basis states $|T\rangle$ can be expressed as
  \begin{equation}
  \label{eqHonT}
      e^{-i dt \hat{H}}|T\rangle=\int D \phi e^{-i dt N\mathcal{H}[T,O]} e^{-N T_{i j} O_{i j}}|\phi\rangle.
  \end{equation}
 where the induced Hamiltonian reads
  \begin{equation}
      \mathcal{H}[T, O]=R\left(-2 T_{i j} O_{j i}+4 T_{i k} O_{k l} T_{l j} O_{j i}\right)+U\left(2 T_{i i}-4 T_{i k} O_{k l} T_{l i}\right)+\lambda O_{i i}^{2}.
  \end{equation}
  
For a general state $|\Psi\rangle$,

  \begin{equation}
  \label{infinite t}
e^{-i d t \hat{H}}|\Psi\rangle= \int D {T}^{(1)}  D {P} D{T}^{(0)}\left|{T}^{(1)}\right\rangle
 e^{N P_{ij}\left(T_{ij}^{(1)}-T_{ij}^{(0)}\right)-i N d t \mathcal{H}\left[{T}^{(0)}, {P}\right]}\Psi(T^{(0)}) ,
\end{equation}
where  $DP=\prod_{i\leq j}dP_{ij}$ along the real axes.
  The integration of $T^{(1)}$ gives $\delta(P_{ij}-O_{ij})$ and the integration of $P$ reproduce Eq.\ref{eqHonT} for $|\Psi\rangle$.
  A finite time evolution gives a path integral as
  
  \begin{equation}
  \label{finite t}
      e^{-i t \hat{H}}|\Psi\rangle=\int \mathcal{D} T \mathcal{D}P \left| {T}(t)\right\rangle e^{i S[{T}, {P}]} \Psi(T(0))
  \end{equation}
  
  where 
  \begin{equation}
      S[T,P]=N\int_{0}^{t} d \tau\left[-i  P_{i j} \partial_{\tau} T_{i j}-\mathcal{H}[T,P] \right].
  \end{equation}
 The evolution preserves the form of the state $|\Psi\rangle$, so that $T$ and $P$ can be taken as classical variable as long as (\ref{eq:ineq}) is satisfied. Their evolution equations are given by the classical path:
    \begin{equation}\label{eom} \begin{aligned} 
-i \partial_{\tau} T_{i j} &=R\left(-2 T_{i j}+8 T_{i k} P_{k l} T_{l j}\right)-4 U T_{i k} T_{k j}+2 \lambda P_{i i} \delta_{i j}, \\
i \partial_{\tau} P_{i j} &=R\left(-2 P_{i j}+8 P_{i k} T_{k l} P_{l j}\right)+U\left(2 \delta_{i j}-4 P_{i k} T_{k j}-4 T_{i k} P_{k j}\right),
\end{aligned}\end{equation}
with initial $P$ and $T$ taking $\bar{P}$ and $\bar{T}$ from initial $|\Psi\rangle$.


In the weak coupling range, $\left|T_{i \neq j}\right| \ll e_{i i}, e_{j j}$ with $T_{i j}=e_{i j}+i t_{i j}$, the entanglement entropy between a subset of sites $A$ and its complement $\bar{A}$ is
\begin{equation}\label{een}
S_{A}=N\left[\sum_{i \in A, j \in \bar{A}}\left(-\ln \frac{\left|T_{i j}\right|^{2}}{4 e_{i i} e_{j j}}+1\right) \frac{\left|T_{i j}\right|^{2}}{4 e_{i i} e_{j j}}+O\left((T / e)^{4}\right)\right]+O\left(N^{0}\right).
\end{equation}

The corresponding mutual information between sites $i$ and $j$ is

\begin{equation}\label{eMI}
I_{i j}=N\left[\left(-\ln \frac{\left|T_{i j}\right|^{2}}{4 e_{i i} e_{j j}}+1\right) \frac{\left|T_{i j}\right|^{2}}{4 e_{i i} e_{j j}}+O\left((T / e)^{4}\right)\right]+O\left(N^{0}\right).\end{equation}

These equations further certify that the collective variables $T_{ij}$ can measure mutual correlations. We use them to compute the entanglement evolution along the classical paths. 

\subsection{Dynamics}

To explore their dynamics, first set $T$ and $P$ to be diagonal at $\tau=0$. The collective variables will remain diagonal during the evolution since all sites are decoupled. A real static solution is $T_{ij}=T_*\delta_{ij}$, $P_{ij}=P_*\delta_{ij} $ with $\left(T_{*}, P_{*}\right)=\left(\frac{1}{2}\left(\frac{\lambda}{U}\right)^{1 / 3}, \frac{1}{2}\left(\frac{U}{\lambda}\right)^{1 / 3}\right)$. 

We perturb this solution by adding neighbouring couplings. One simple choice is a 1D chain of period $L$.  Denoting $[i-j]_L=min\left(|i-j|,L-|i-j|\right)$ as the distance between site $i$ and $j$, it reads
 \begin{equation}  \begin{aligned} \label{setup1}
& T_{ij}(0)=T_*\delta_{ij}+\varepsilon\delta_{[i-j]_L,1}\\
& P_{ij}(0)=P_*\delta_{ij}.\\
\end{aligned}\end{equation}
where $\varepsilon$ is a constant that controls the initial coupling strength.

\begin{figure}[tbp]

\begin{subfigure}{.5\textwidth}
  \centering
  \includegraphics[width=.9\linewidth]{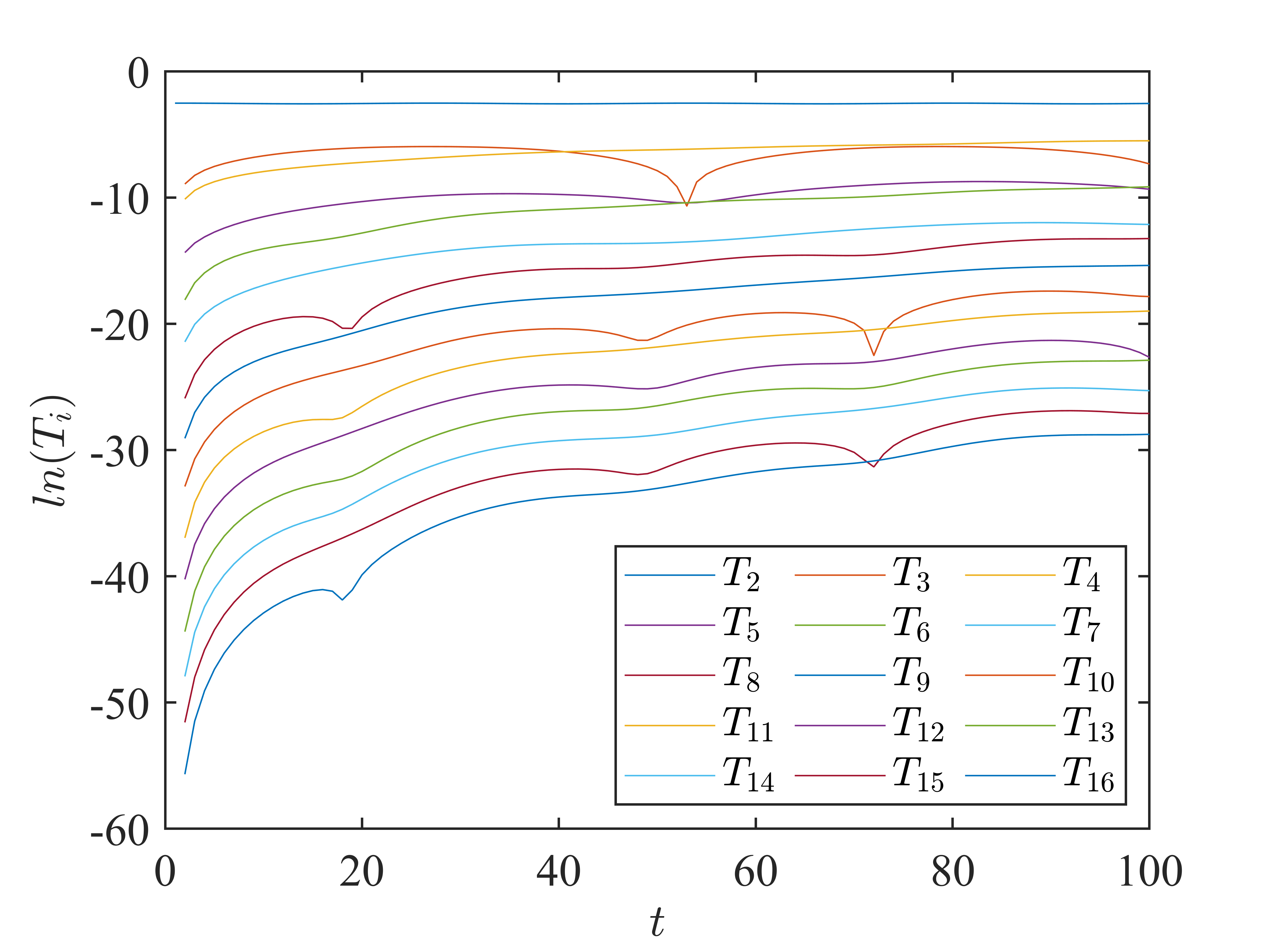}  
  \caption{}
  \label{f_tearly}
\end{subfigure}
\begin{subfigure}{.5\textwidth}
  \centering
  \includegraphics[width=.9\linewidth]{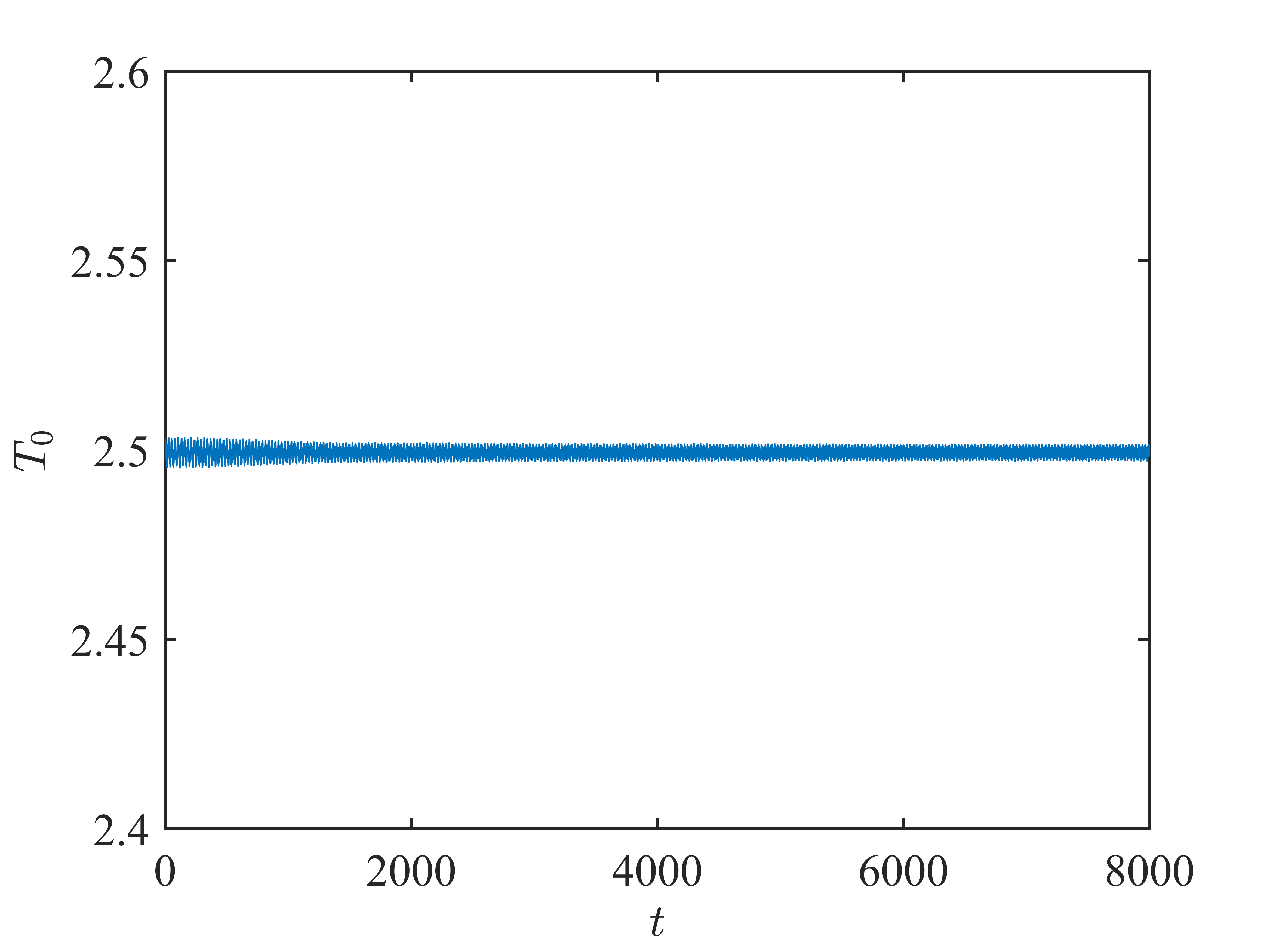}  
  \caption{}
  \label{f_t0}
\end{subfigure}
\begin{subfigure}{.5\textwidth}
  \centering
  \includegraphics[width=.9\linewidth]{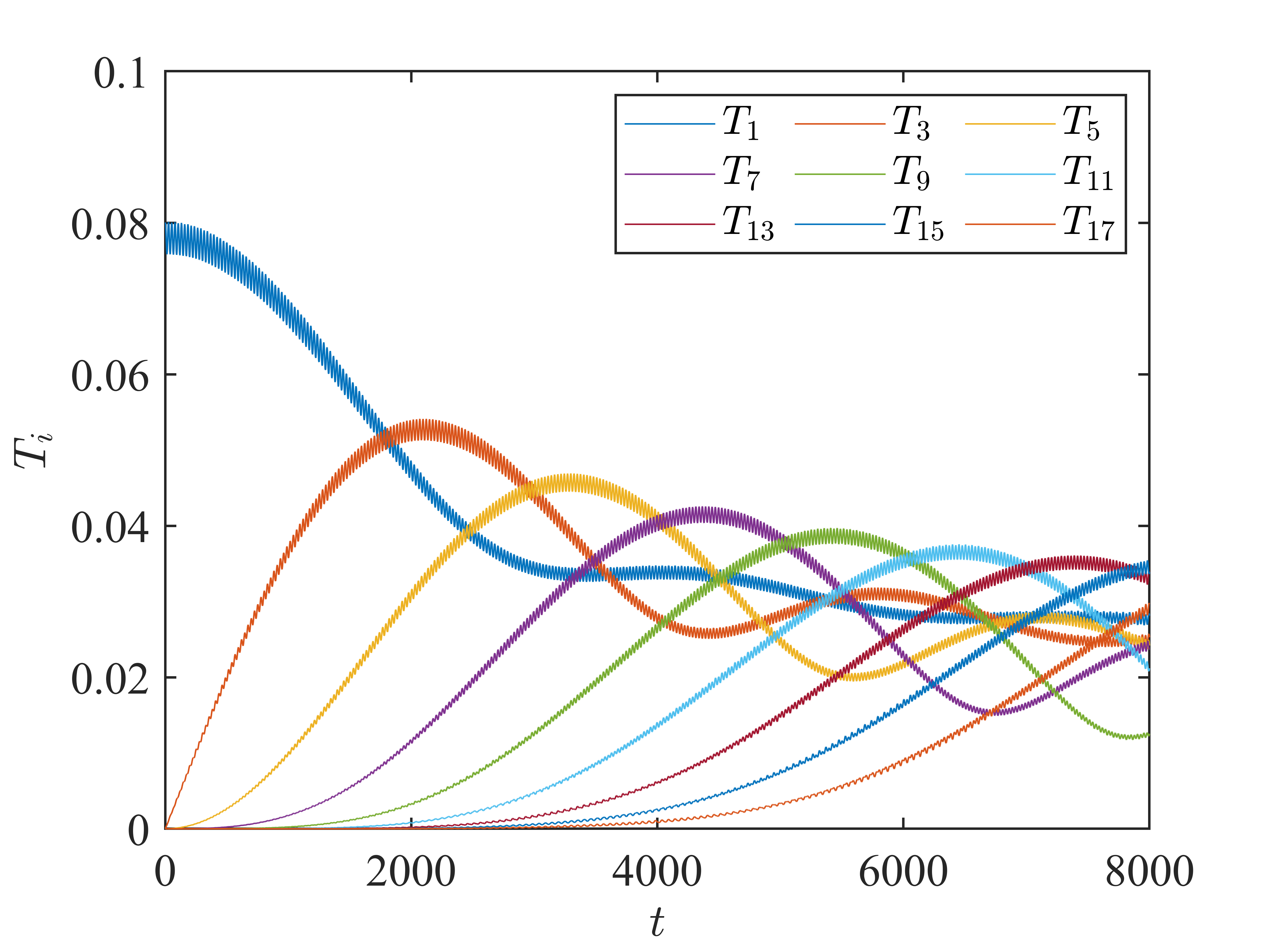}  
  \caption{}
  \label{f_todd}
\end{subfigure}\begin{subfigure}{.5\textwidth}
  \centering
  \includegraphics[width=.9\linewidth]{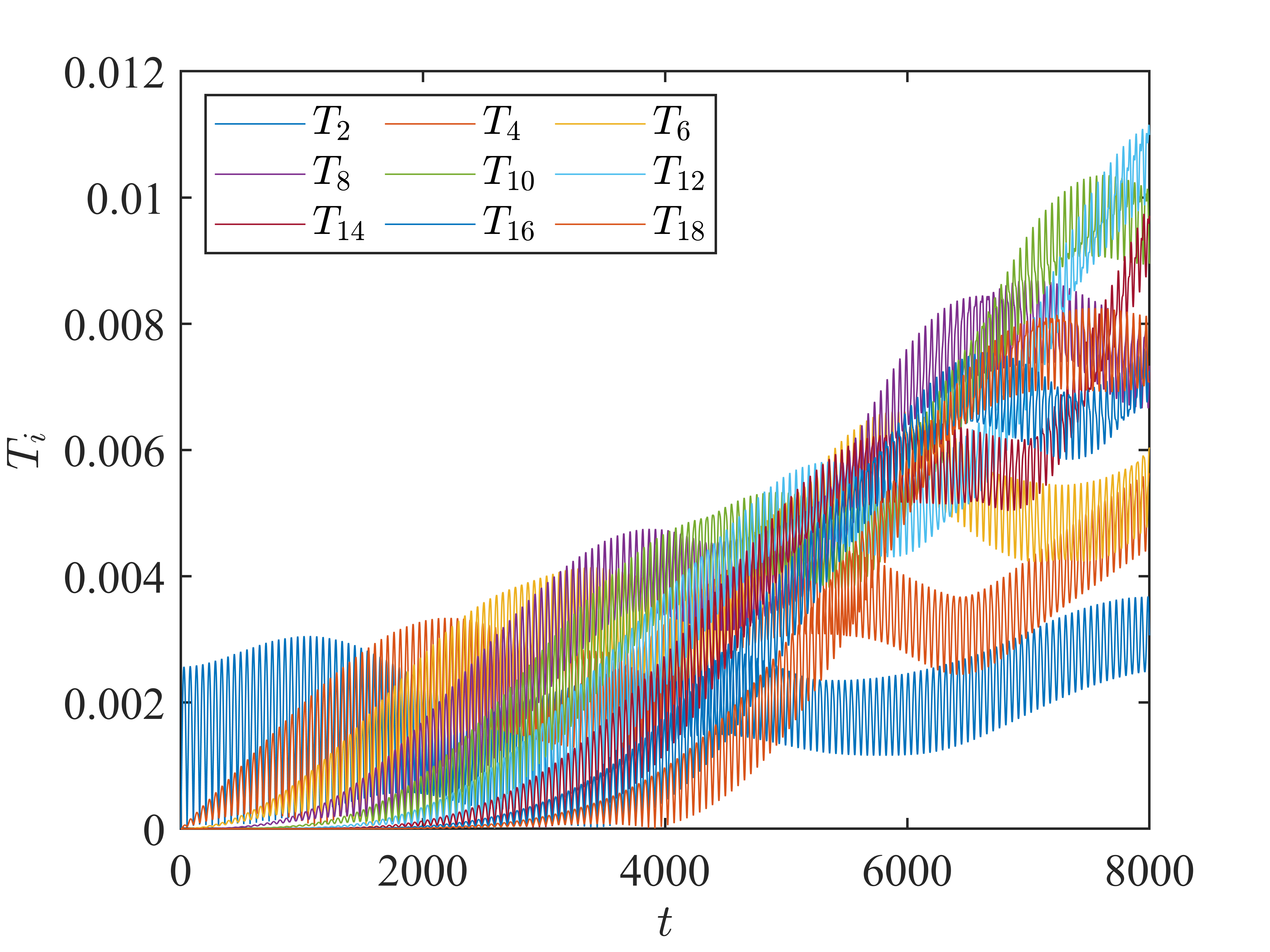}  
  \caption{}
  \label{f_teven}
\end{subfigure}
\caption{Evolution of $|T_i|$.  (a) is the early dynamics of $|T_i|$ in log scale. (b) plots the stable $|T_0|$. (c) and (d) plot separately the dynamics of some odd and even $|T_i|$.}

\label{Ti}
\end{figure}

Such a state will evolve to create wider correlations. We solve for the $T_{ij}(t)$ and $P_{ij}(t)$ numerically, as detailed in the appendix of \cite{Lee:2018hsd}. Because of the symmetries in Eq.\ref{eom} and Eq.\ref{setup1}, $T_{ij}$ only depends on the relative distance $[i-j]_L$. From now we denote $T_{ij}$ by the relative distance as $T_{[i-j]_L}$. Some of their absolute values are plotted in figure \ref{Ti} \protect\footnote{$L=81, U=R=1, \lambda=125$ and $\varepsilon=0.08$ are assumed unless specified. Tuning either the constants or the initial  correlation $\varepsilon$ can controls the spreading speed of collective variables. We choose a set of coefficients which provides a moderate speed for our choice of $L$}. $T_0$ stays around initial value $T_*$ with perturbation of order $\varepsilon^2$. The early $T_i$ is exponentially suppressed over distance. $T_1$ drops from $\varepsilon$ and stabilizes around a smaller value. The others start at zero. Among them, the odd terms grow to descending peaks of order $\varepsilon$ site by site, and fall to oscillate around a smaller value.  The even terms are unstably growing to magnitude of order $\varepsilon^2$. The dynamics of $T_{ij}$ depicts a wave propagating in odd terms at seemingly constant speed of order $\varepsilon^2$. This wave will be our major topic in the next section. Appendix \ref{chap_appendix} gives a phenomenological explanation for the propagation of the waves. There we show how the former peaks combine into resonant oscillations that arouse later peaks. Several signature quantities are argued to be of scaling forms in $\varepsilon$. 

In this paper, we try to demonstrate within these setups, that the notion of "locality" is approximately preserved in some finite time. In section \ref{chap_entanglement}, we continue the discussion of propagating wave in the language of quantum information theory. In section \ref{chap_entanglement}, we will study the quantum butterfly effects and the effective light-cones of this model.

\section{Entanglement spreading}
\label{chap_entanglement}
\subsection{Definitions}

\begin{figure}[tbp]
\begin{subfigure}{.5\textwidth}
  \centering
  \includegraphics[width=.9\linewidth]{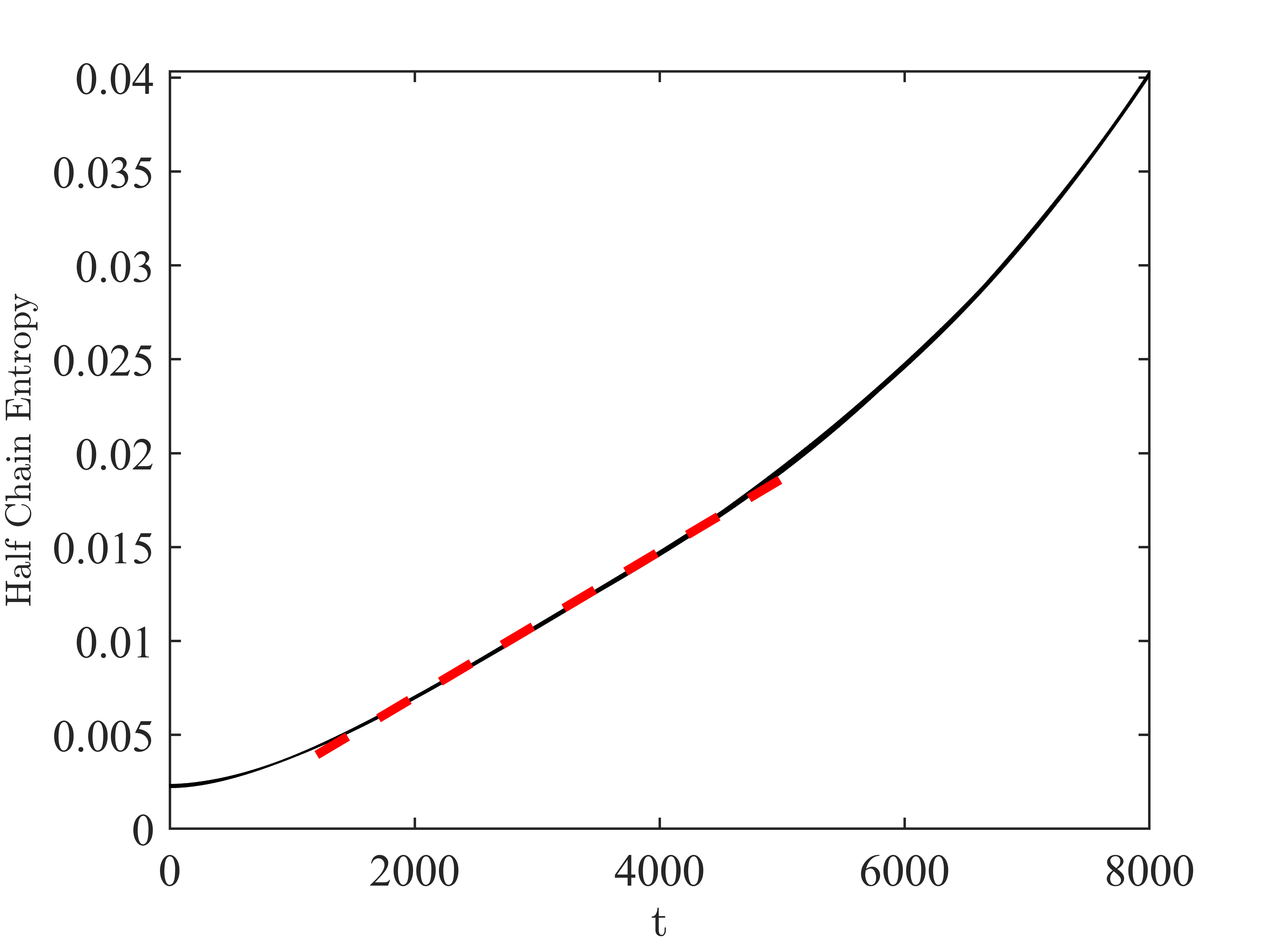}  
  \caption{}
  \label{f_s}
\end{subfigure}
\begin{subfigure}{.5\textwidth}
  \centering
  \includegraphics[width=.9\linewidth]{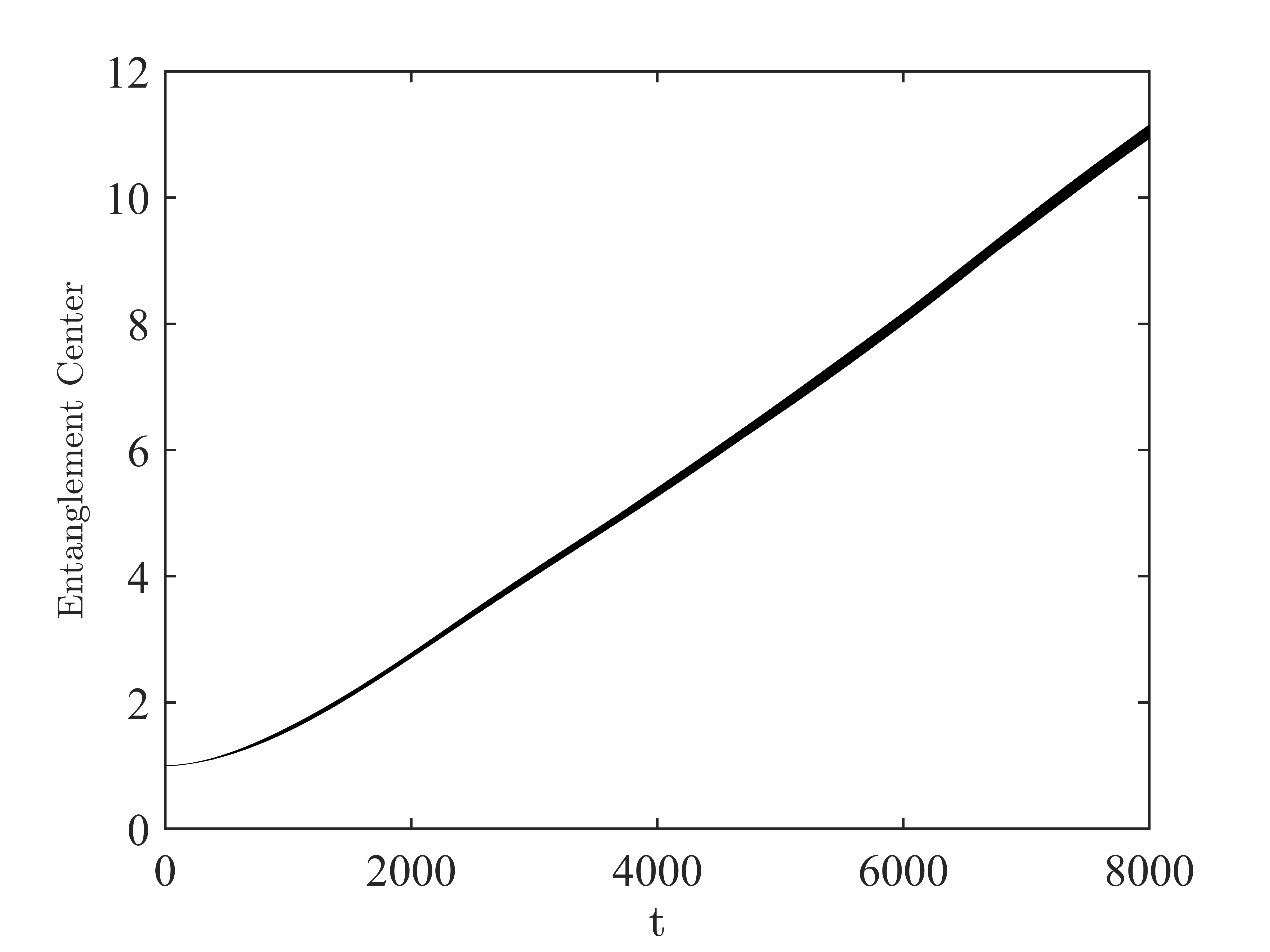}  
  \caption{}
  \label{f_x}
\end{subfigure}
\caption{(a) is the dynamics of half chain entanglement entropy. The red line denotes a linear growth stage. (b) is the dynamics of entanglement center.}
\label{f_ent}
\end{figure}

In the study of quantum information, the creation and spreading of entanglement is a central problem for a quenched quantum system. It is found in both holographic and spin systems, the entanglement of a subregion seems to grow at a constant rate before approaching local equilibrium\cite{tsunami,cardy}. If we look at a series of subregions, the saturation time increases linearly with region size. In 1D systems, this “propagation” of equilibrium can be understood in terms of "information quasi-particles". They transport quantum information uniformly across the system, bringing a linear growth of entanglement, and saturating a subregion when the first quasi-particle has gone through the subregion. We can directly see these quasi-particles in spin systems by monitoring two-point correlations\cite{correlationspreads}. The speed of either the entanglement or the quasiparticles is a constant that depends on the system under local interactions. 

The entanglement dynamics with long-range interaction is still an on-going topic. The most studied cases are systems with power-law interactions \cite{longrange,longrangeentanglementcone,longrangemagcone,differentregime}. By tuning the exponent, the system is believed to see different regimes of dynamics, from (quasi-)linear entanglement growth that is similar to a local theory, to logarithmic entanglement growth with possibly divergent quasiparticle velocity. These are closely related to the topic of the next section.

In this section, we focus on the entanglement entropy description of the correlation spreading. By the word "correlation", we mean both the entanglement between sites, and the universal coupling of the Hamiltonian, as they are designed to be related. We will give our definitions of three characteristic velocities to describe the entanglement dynamics and their dependence on the initial states.

\begin{figure}[tbp]
    \centering
    \includegraphics[width=.6\textwidth]{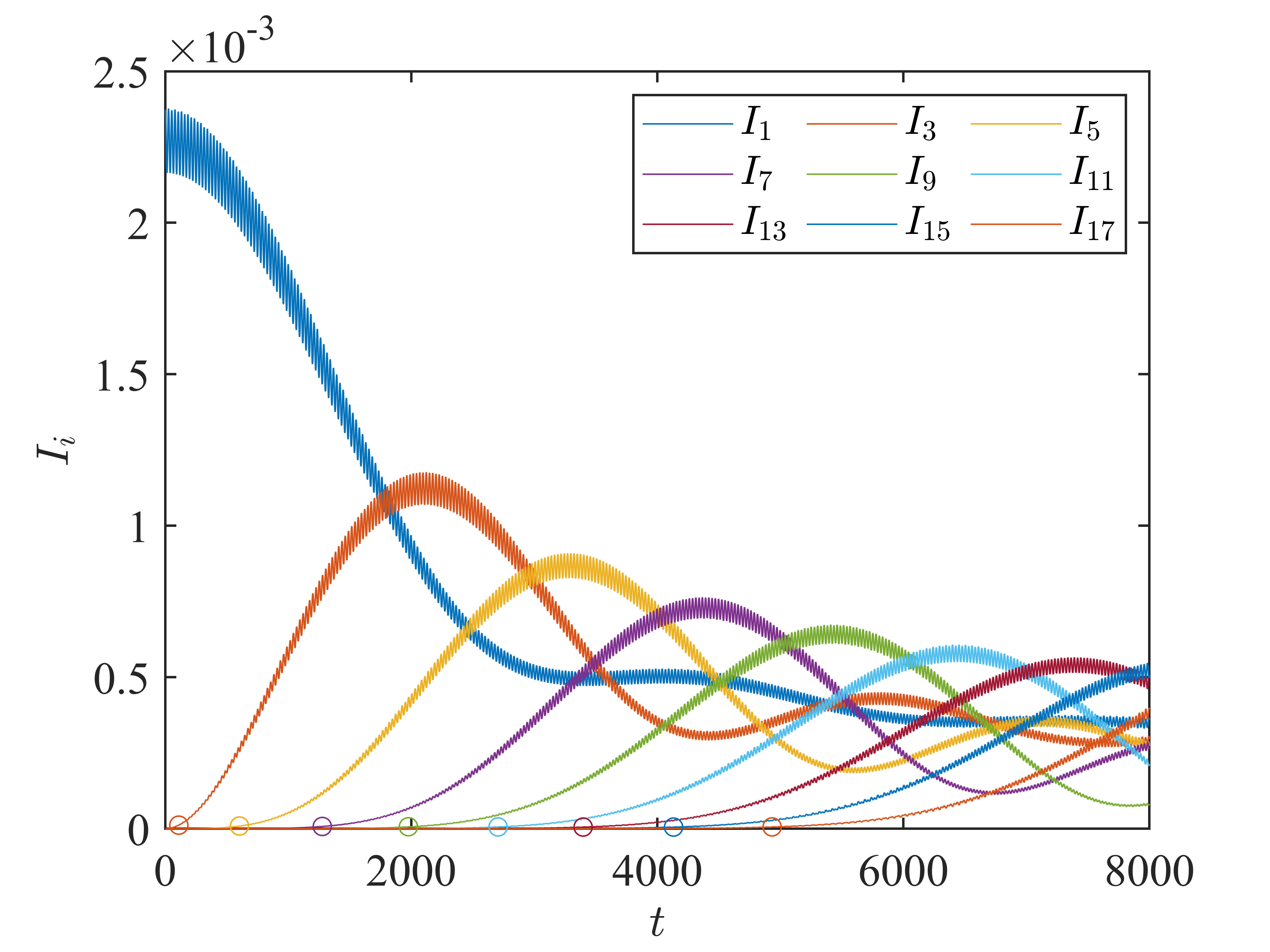}
    \caption{Dynamics of $I_i$. The circles denote the instigation points, truncated at 1 percentage of the peak values.}
    \label{f_i}
\end{figure}

As in Eq.\ref{eMI}, the mutual information $I_{ij}$ between site $i$ and $j$ has similar evolution as $T_{ij}$. $I_{[i-j]_L}$ are plotted to first order of $N$ in figure \ref{f_i}\footnote{We omit the factor $N$.}. The quantum information of local sites leaks out through the propagating wave packets. We define the speed of the wave peaks as the information speed $V_I$. The peak values decrease as the waves go forward. It can be justified by the fact that some information stays in the residue——the $I_{i}$s do not fall to zero after the peak.
Meanwhile, the wavefront is broadening with time. We can notice this by the increasing lag of the peak from the "instigation point", which refers to the position of the curve where the oscillation begins to rise. These instigation points of each site forms an information light-cone, outside which little information is leaked from the origin. The corresponding slope is defined as the information light cone speed $V_{ILC}$. Numerically, we truncate at 1 percent of the peak value as the instigation. The clear difference of the $V_{ILC}$ and the $V_I$ suggests a rich structure inside the propagating wavefront. Detailed discussions of the light cones are in the next section.

The other signature quantity is the entanglement entropy. As in Eq.\ref{een}, it is a naive sum of all relevant site-to-site mutual information, which enables a natural quasi-particle description of the entropy evolution. These particles start off at each site and enter a certain region with the same velocity, which brings a steady increase of the entanglement entropy. Figure \ref{f_s} is the entanglement entropy of one half of the system. It deviates from linearity because the residue of the peaks abnormally increases the entropy. In addition, the entropy follows neither area law nor volume law during the evolution. The usual definition of entanglement speed does not rely on the choice of a specific subregion. Without loss of generality, we define the entanglement center as
\begin{equation}
    X  =\frac{S_{\text{Half Chain}}}{S_{\text{Single Site}}}.
\end{equation}
It is plotted in figure \ref{f_x}. For odd $L$, $X=\frac{\sum_{2=<j<L/2+1}(j-1)I_{1j}}{\sum_{2<=j<L/2+1}I_{1j}}$, giving the weighted sum of distance travelled by the  "quasi-particles". For even $L$, it 
deviates little from such a form. $X$ can describe the "center of mass" for quantum information quasi-particles. It is natural to define its speed as entanglement center speed, $V_{CE}$.

\subsection{State dependence}

\begin{figure}[tbp]
    \centering
    \includegraphics[width=.6\textwidth]{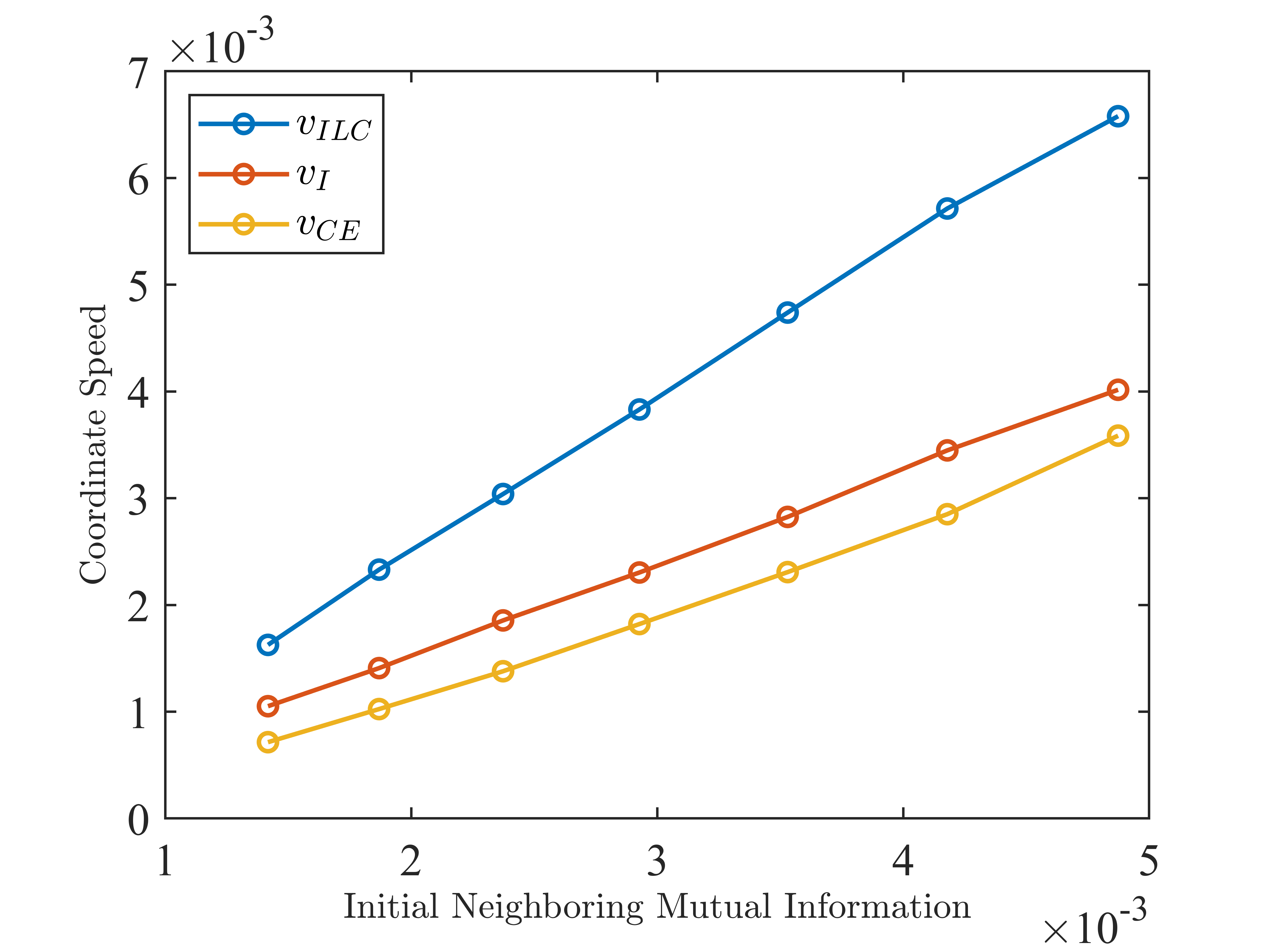}
    \caption{$V_I$, $V_{ILC}$ and $V_{CE}$ against different initial neighboring mutual information.}
    \label{f_speed}
\end{figure}

Keeping the same Hamiltonian, we change only the initial neighboring mutual information by $\varepsilon$ and plot $V_I$, $V_{ILC}$ and $V_{CE}$ in figure \ref{f_speed}. It is clear that $V_{ILC}$ is greater than $V_I$. The wave packets are broadening with time, i.e. the quasi-particle are gradually delocalizing. It is closely related to the light cone behavior of Sungsik’s model discussed in the next section, that the light cone velocity is greater than the butterfly velocity. $V_I$ is slightly higher than $V_{CE}$, suggesting that the wave packet is not symmetric about the peak. It is close to zero at the instigation points but non-zero at the residue side.  The entanglement center thus lags behind the peak. As showed in figure \ref{f_speed}, three speeds are all proportional to initial neighboring mutual information. It meets with physical intuitions that higher energy leads to higher speed. We show in the appendix that the coordinate speed of the wave is of order $\varepsilon^2$, while the initial mutual information is proportional to $\varepsilon^2$ for small $\varepsilon$. This result suggests a linear relation between coordinate speeds and the initial mutual information.   \cite{brian} argues in general that the initial entanglement density has strong influence on the information velocity. It is however more straightforward in the Sungsik model where the coupling depends on the correlations.

\section{Emergent light cones}
\label{chap_lightcone}
\subsection{Operator spreading}

Relativistic theories are equipped with exactly vanishing space-like commutators. Such rigid light-cones ensure that causality is respected in all inertial frames. 
In non-relativistic quantum mechanics, for normalized local operator $\mathcal{O}_y$ located at $y$ and the Heisenberg picture of another normalized local operator $\mathcal{O}_x(t)=e^{i\mathcal{H}t}\mathcal{O}_xe^{-i\mathcal{H}t}$, the Lieb-Robinson bound declares that  $|[\mathcal{O}_x(t),\mathcal{O}_y]| \leq k e^{-\lambda(|x-y|-v_Lt)}$.  Here $k$ and $\lambda$ are constants determined by the system. $| \cdot |$ denotes taking norm. The commutator is exponentially small for $|x|>v_Lt$, which defines an effective light-cone with the  Lieb-Robinson speed $v_L$. The  out-of-time-order correlators (OTOC) are usually defined as $F(x-y,t)=\langle |[\mathcal{O}_x(t),\mathcal{O}_y]|^2\rangle_\rho$ \footnote{In some literature, OTOC is defined as $F(x-y,t)=\langle |\mathcal{O}_x(t)\mathcal{O}_y\mathcal{O}_x(t)\mathcal{O}_y\rangle_\rho$. For Hermitian operators, $F(x,t)=2-2C(x,t)$.}, which takes the ensemble average of the squared commutators. Its deviation from zero can quantify how local operators evolve to overlap with distant ones.  In most chaotic systems, it is believed that the region where $F(x,t)$ has grown to $O(1)$ expands with constant velocity \cite{localizedshocks}. Such ballistic spreading of local operators is named quantum butterfly effect after the classical chaos theory. The early behavior of the OTOC is argued to be $F(x,t)\sim e^{-\lambda'(x-v_Bt)}$ \cite{earlyotoc}. It saturates an ensemble-dependent Lieb-Robinson bound with butterfly velocity $v_B$, which should be naturally not greater than than a universal effective light-cone velocity. The propagating front, where the OTOC is between 0 and $O(1)$, may broaden with time. We have already seen similar effects in the quasi-particle description, where $V_{ILC}$ is higher than $V_I$ and leads to a broadening front of the information wave. The diffusion of the butterfly front is well studied in random circuits. The width is believed to follow universal scaling forms \cite{scalingformofbutterfly} and governed by some hydrodynamics equations \cite{hydro1,hydro2}, thus given the name hydrodynamical effects.

In this section, we will demonstrate how operators spread and form light-cones that are both emergent and effective. By "emergent" we mean its shape and velocity depends on the states of the system. The word "effective" comes from the spirit of Lieb-Robinson bound that a non-relativistic quantum theory can have a light-cone with exponentially small space-like commutators. Different levels of couplings give rise to a rich internal structure of the butterfly front. The increasing long-range interactions slowly melt the light-cones.

Due to the semi-classical nature of the equation of motion \ref{eom}, we can adopt an easy method to approximate commutators: to disturb a local variable and see how it affects the evolution of distant variables. We shift the value $T_{ii}$ of a state $|\Psi\rangle$  by acting $|\Psi'\rangle=e^{i\alpha O_{ii}}|\Psi\rangle$. For a chosen local operator $\mathcal{O}_j$ and a small real $\alpha$, $\langle\Psi'|\mathcal{O}(t)|\Psi'\rangle-\langle\Psi|\mathcal{O}(t)|\Psi\rangle= \langle\Psi|e^{-i\alpha O_{ii}}e^{i t \hat{H}}\mathcal{O}e^{-i t \hat{H}}e^{i\alpha O_{ii}}|\Psi\rangle-\langle\Psi|e^{i t \hat{H}}\mathcal{O}e^{-i t \hat{H}}|\Psi\rangle= i\alpha\langle\Psi|[\mathcal{O}(t),O_{ii}]|\Psi\rangle+O(\alpha)$. Thus we are actually calculating expectation values of commutators using collective variables.

Numerically, we choose a state $|\Psi\rangle$ and shift one of $T_{ii}$ by a small  $i\alpha$ to $|\Psi'\rangle$. They evolve to give $T_{jj}$ and $T'_{jj}$. $C(j,t)=\left|\frac{T'_{jj}(t)-T_{jj}(t)}{\alpha T_{jj}(t)}\right|$ for each $j$ gives approximation of normalized commutators.

\subsection{Emergent light cone}
\begin{figure}[tbp]
\begin{subfigure}{.5\textwidth}
  \centering
  \includegraphics[width=.9\linewidth]{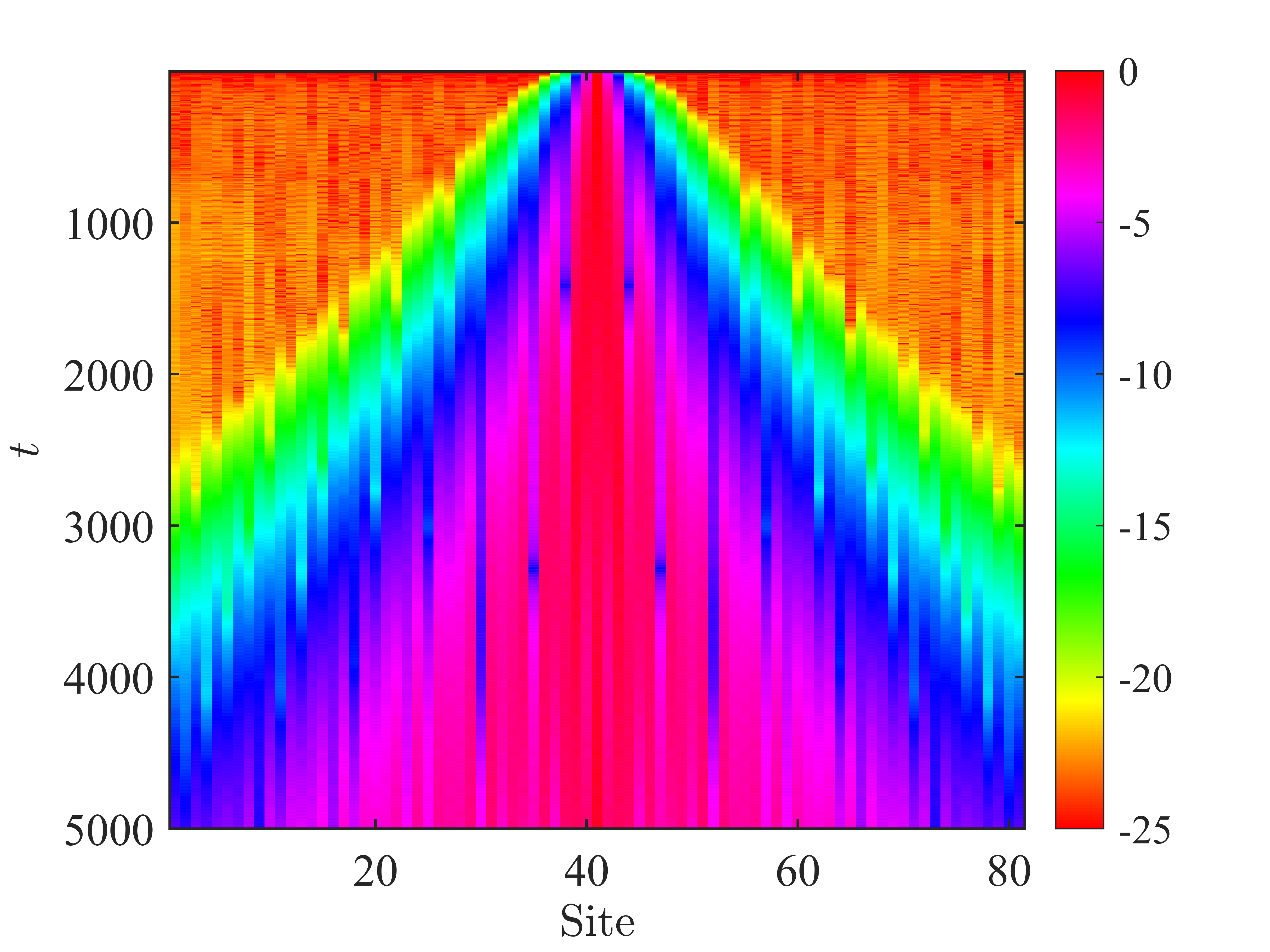}  
  \caption{}
  \label{lc0}
\end{subfigure}
\begin{subfigure}{.5\textwidth}
  \centering
  \includegraphics[width=.9\linewidth]{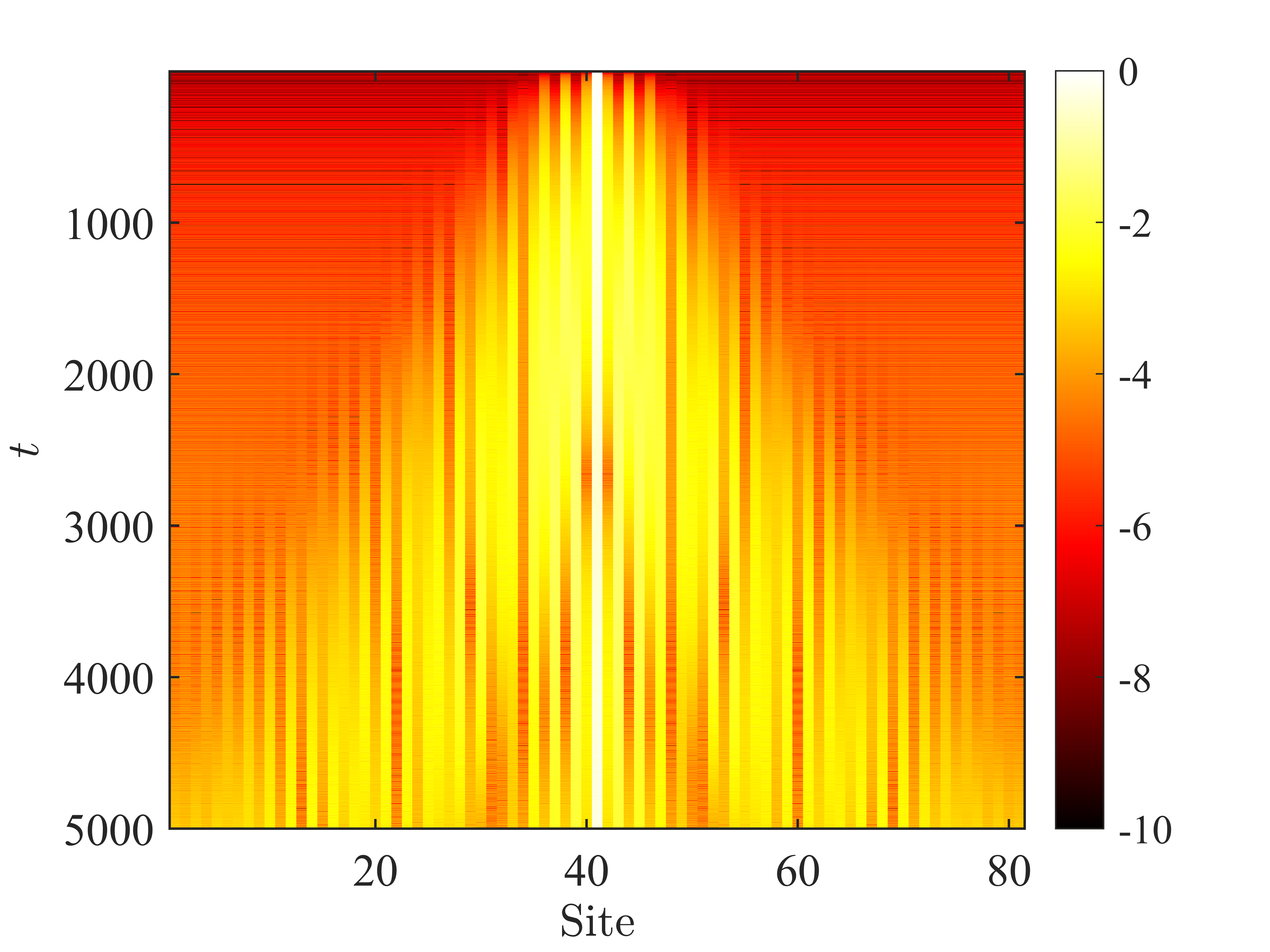}  
  \caption{}
  \label{lc1}
\end{subfigure}
\begin{subfigure}{.5\textwidth}
  \centering
  \includegraphics[width=.9\linewidth]{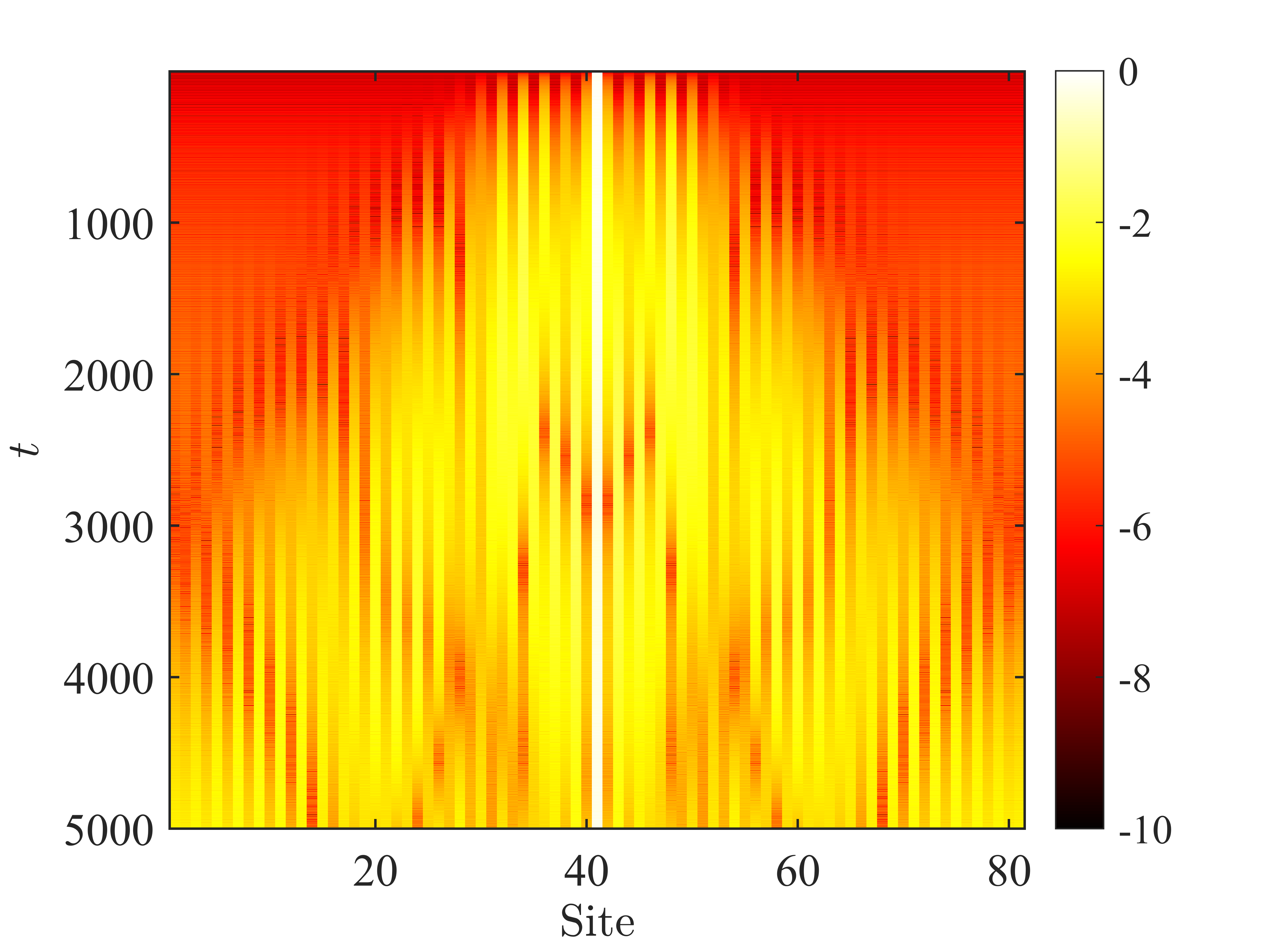}  
  \caption{}
  \label{lc2}
\end{subfigure}
\begin{subfigure}{.5\textwidth}
  \centering
  \includegraphics[width=.9\linewidth]{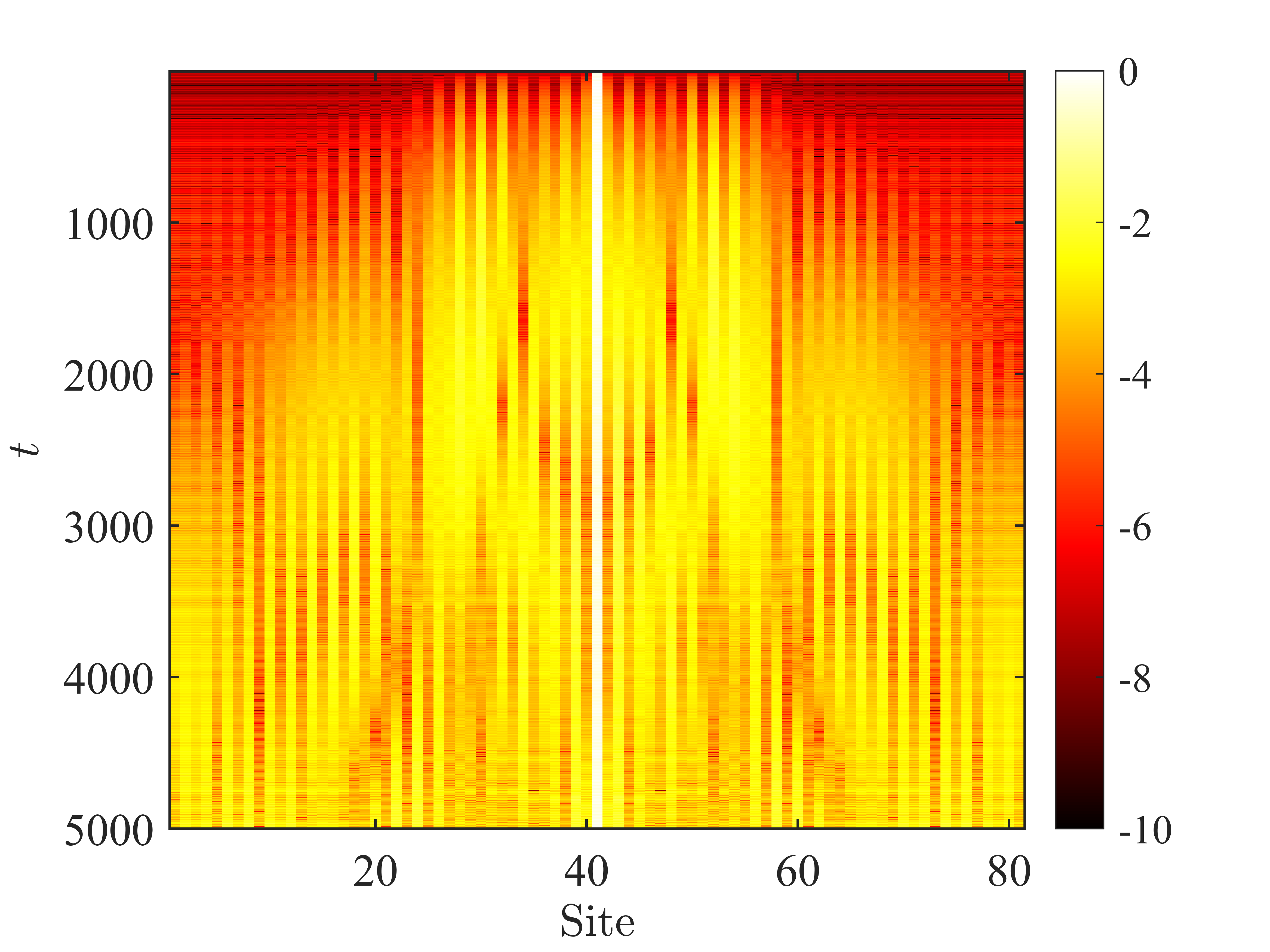}  
  \caption{}
  \label{lc3}
\end{subfigure}
\caption{Density plot of $\ln{ C(i,t)}$ vs site i and time t. We set states at $t'=0,2000,4000,6000$ in figure \ref{Ti} with $T_{41,41}$ perturbed to $T_{41,41}+0.0001i$ as initial states in (a), (b) (c) and (d).}
\label{lc}
\end{figure}

The diagrams of the operator spreading, namely $C(i,t)$, can reflect the locality structure of the perturbed states $|\Psi\rangle$. In order to see how it changes during the evolution, the perturbed states are chosen to be the states that have evolved from Eq.\ref{setup1} for a certain time $t'=0,2000,4000,6000$, as in the states in figure \ref{Ti}. We perturb their $T_{41,41}$ to $T_{41,41}+0.0001i$ and plot $\ln{ C(i,t)}$ separately for each $t'$ in figure \ref{lc}. 

In the figure \ref{lc0}, we just perturb the initial states as in Eq.\ref{setup1}. The correlation strength is exponentially suppressed in the early time. As we have seen in the last section,  higher correlation strength will lead to higher amount of propagating quantum information. We would expect similar thing for the butterfly effects. There are multiple layers of $C(i,t)$ in figure \ref{lc0}, with varying truncation values and slopes. The exponentially small but existing long-range interactions lead to an early arrival of the operators with small magnitudes. The near-range interactions provide slower but stronger spreading of operators that dominates the butterfly effects. As shown in figure \ref{lc0}, the effective light-cone for different choices of the truncation value is different. However, any sufficiently small enough truncation value would provide an equally good effective light-cone, outside which the commutators are negligible. An $O(1)$ truncation qualifies a butterfly cone, for instance the red-pink line. We can again define butterfly velocity and light cone velocity by artificially giving a truncation value. They are still proportional to initial mutual information, which we do not bother to show explicitly. As in the figure \ref{lc0}, a higher truncation has higher velocity. The gap between any two cones are expanding, which can be seen as a generalized hydrodynamical effect.

Notice in $C(i,t)$, we are not able to do an ensemble average, but only sandwich the commutator with a chosen $|\Psi\rangle$. The locality structures are all state dependent. It is widely believed that the butterfly velocity is an ensemble-dependent light-cone velocity \cite{localizedshocks,brian}, which is naturally smaller than the actual light velocity.  In 1D random circuit models, the front size scales as $t^{\frac{1}{2}}$ \cite{scalingformofbutterfly}, so the velocity of either side of its front is asymptotically the same. In figure \ref{lc0}, the gap scales almost linearly in time. We thus give separate names to the state dependent light-cones velocity and the butterfly velocity.

As we are observing the spreading of operators, the background coupling strength is growing. The velocity for each cone thus gets slightly higher. Apart from such bending, the cones are basically linear.  As we have briefly discussed in the last section, the different structures of long-range interaction will lead to different regimes of dynamics. Our results align with general beliefs that a quantum system with exponentially suppressed long-range interaction is likely to have a linear light cone, which corresponds to the linear regime of the power law interaction with large exponent\footnote{The interaction strength of site $i$ and $j$ in such a model is proportional to $1/|i-j|^\alpha$, with $\alpha$ a constant exponent. } \cite{differentregime,protectedlocality}.

\cite{chaosprotectedlocality} showed that, in a variant of the Sachdev-Ye-Kitaev model with non-local interactions, the locality structure is almost preserved for "simple signals" that is carried by quasi-particles. It is similar to our results that the dominant behaviour of the OTOC or the entanglement entropy is linear and controlled by the near-range interactions, while some remnant of non-locality is leaked out through long-range interactions. We conjecture that for non-local models with similar hierarchy of interactions——that the coupling strength is at least inversely suppressed over distance——the information flow is dominated by near-range couplings and thus to some degrees preserves locality. 

In figure \ref{lc1}, the perturbed initial states are the same as the states in figure \ref{Ti} at $t'=2000$. The quasi-particle front has reached $T_7$. Inside this front, the odd coupling strength are of the same magnitude. This reflects in the step-like cones. A group of nearly seven sites form each step. It is clearer in the figure \ref{lc2} and \ref{lc3} where each step is wider, since the perturbed states has wider correlations of order $\varepsilon$. The butterfly velocity thus increases with our choice of $t'$.

From the perspective of the initial 1d geometry of this periodic chain, we can interpret the evolution as a dynamical evolution of the couplings. The growing range of the coupling speeds up the spreading of the operators. It agrees with \cite{nonlocalspeeding} that a higher level of non-locality leads to higher quantum information spreading speeds. After the quasi-particle front reaches most of the system, an originally local operator spreads to the whole system in a very short time. The causality structure gradually melts.

Alternatively, we can interpret the evolution under a changing geometry of the system. Inside the range where the coupling strength are of the same magnitude, we assume the sites to be physically close. This range expands linearly, as suggested by the $T_i$ waves. "Light" can propagate across the system within less time. The system thus shrinks with the inverse of the time, until the whole system collapses to one single pot of glue -- where every site correlates equally strongly with every other site. The notion of locality is completely lost at sufficiently late times.  The notion of locality, and thus also of dimensionality of space is dynamically varying according to the correlations. 

\cite{brian} argues that the butterfly velocity should bound the information velocity, since the former is a maximum velocity of operator spreading for the whole ensemble. It is however not the case here where the butterfly velocity is state dependent. If we regard the butterfly front in figure \ref{lc0} as the one describing the geometry of the dynamics of information quasiparticles, its butterfly velocity is greater than the information velocity.

\section{Discussion}

We study a specific class of non-local models such that the notion of locality is dictated by the entanglement of the initial states.
We would like to find out if models with only an emergent and state dependent sense of locality satisfies similar properties as more conventional local models. 
We explored the problem from two different perspectives -- the spreading of quantum information from the computation of mutual information between different sites, and the growth of the effective size of Heisenberg operators under dynamical evolution. We find that from either perspective, there is an emergent light-cone, and the notion of entanglement speed/ butterfly speed that characterizes these light cones exist. However, the relations observed between them in more conventional local models do not appear to be satisfied. 

Moreover, since locality depends on the state, as a state evolves from given initial conditions that supplied the locality structure, the notion of locality is gradually lost as time evolution effected by an underlying non-local Hamiltonian progresses.  Recall the pattern of dynamical evolution of entanglement in figure \ref{f_i} and the growth of the Heisenberg operators shown in figure \ref{lc0}, where both display a relatively sharp linear light-cone reflecting the exponentially suppressed initial interactions. In this regime, the behaviour of the model is not different from quasi-local models whose range of interaction decays only as  power laws with large exponent \cite{differentregime,improvedLB}.  As explained in the appendix, the oscillation drove wider correlations, and the increasing range of the interactions would destroy the light cones. Typically, the breakdown of light-cones are seen by directly tuning the coupling strength in the Hamiltonian\cite{longrangeentanglementcone,longrangemagcone,differentregime}. But in our case, the couplings depend directly on the correlations, which itself is evolving. As a result, the light-cones slowly melt under a single fixed Hamiltonian. 

Along with similar observations from non-local spin models\cite{chaosprotectedlocality,differentregime}, we conjecture that the hierarchy of the interaction over different distances provides different ``layers'' of light-cones. The short range interactions lead to linear cones, which can be explained by quasi-particles, while longer range interactions lead to quasi-linear or even logarithmic shaped cones. When long range interactions are sufficiently suppressed, the effective light cones are dominated by linear behaviour with little remnant of non-locality. When the long range interactions dominate over the near range interactions, the linear light cones break down. At an intermediate stage such as that depicted in  Fig.\ref{lc}, the light cone structure is enriched and the multiple layers are visible at the same time.

These would potentially be interesting diagnostics of models that are genuinely local or only local in a state-dependent way -- which have important implications particularly in the search for signatures of quantum gravity -- which is believed to be intrinsically non-local due to requirements of diffeomorphism invariance.

It would be interesting to explore if there are universal rules governing the speed of information spread vs operator growth in these non-local models. This would be left for future investigations.

\acknowledgments

LYH acknowledges the support of NSFC (Grant No. 11922502, 11875111) and the Shanghai
Municipal Science and Technology Major Project (Shanghai Grant No. 2019SHZDZX01),
and Perimeter Institute for hospitality as a part of the Emmy Noether Fellowship programme. KJ acknowledges the support of Fudan Undergraduate Research Opportunities Program.

\appendix
\section{Phenomenological analysis of information waves}
\label{chap_appendix}

This section is partially based on the ideas in \cite{lee2018emergent}. Denote $V_{s}=\begin{pmatrix}T_{s}\\P_{s}\end{pmatrix}$. Numerical results have shown that under our choice of constants: $|V_{s>0}|$ is not greater than $\varepsilon$; $|V_0|$ stays near initial value $V_*\equiv \begin{pmatrix}T_*\\P_*\end{pmatrix}$. Assume generally that $\varepsilon$ is small and $V_*$ is $O(1)$.  Rewrite Eq.\ref{eom} as 
\begin{equation}
        \partial_tV_{s}=i M_{s}V_{s}+i A_{s}
    \end{equation}
where $M_{s}$ contains all order 1 coefficients of $V_s$,

\begin{equation} \label{M}
        M_{s}= \left(\begin{array}{cc}-2 R+16 R T_0P_0-8 U T_{0} & 8 R T_{0}^{2}+2 \lambda \delta_{s,0} \\ -8 R P_{0}^{2}+8 U P_{0} & 2 R-16 R T_{0} P_{0}+8 U T_{0}\end{array}\right).
    \end{equation}

The rest terms are combined into $A_s$.
$\partial_tV_{0}=i M_{0}V_{0}$ is the ultra local equation of motion, which gives the static solution $V_*$. Because $A_0$ are at best of order $\varepsilon^2$, $V_0$ are only slightly perturbed away from the initial value $V_*$ as expected.

For $s>0$, $M_{s}$ is a constant matrix. Each individual product of collective variables in $A_s$ contain at least two $V_{s>0}$. Thus they are all at best of order $\varepsilon^2$. $\partial_tV_{s}=i M_{s}V_{s}+i A_{s}$ are resonant complex oscillation systems with different initial condition and small force term $A_s$. We can express the solution as: $V_s(t)=V_{s,+}(t)+V_{s,-}(t)$ where $V_{s,\pm}=\alpha_{s,\pm}(t)e^{\pm i\omega t}v_\pm$. $\pm\omega = \pm2 \sqrt{2 U^{2 / 3} \lambda^{1 / 3}\left(2 U^{2 / 3} \lambda^{1 / 3}-R\right)}$ and $v_\pm^T=\left(\frac{R U^{2 / 3} \lambda^{2 / 3}-2 U^{4 / 3} \lambda \pm \sqrt{2 U^{2} \lambda^{5 / 3}\left(2 U^{2 / 3} \lambda^{1 / 3}-R\right)}}{-R U^{4 / 3}+2 U^{2} \lambda^{1 / 3}}, 1\right)$ are the eigenvalues and the eigen-vectors of the mass matrix. $\alpha_{s,\pm}$ are their complex amplitudes.  Note that $\pm\omega$ can be understood as the natural frequency of the oscillators. 

It suits our numerical results amazingly well as we can extract steady amplitudes $\alpha_{s,\pm}$ from $T_s$ and $P_s$, given the relation of the $v^1_{\pm}$ and $v^2_{\pm}$. Because all $V_{s>0}$ share the same frequency, $A_s$ have many small terms with frequency matching the resonance frequency $\omega$, which could drive the oscillation amplitudes. Odd order of $v$ gives resonance driving/damping force such as $V^1_{a,+}V^2_{b,-}V^1_{c,+}=\alpha_{a,+}\alpha_{b,-}\alpha_{c,+}e^{i\omega t}$. They appear in $T_aP_bT_c$ or $P_aT_bP_c$ terms in $A_s$ for $|a\pm b\pm c \pm s|=0$ or $L$, because the indices of $T_{ij}P_{jk}T_{kl}$ or $P_{ij}T_{jk}P_{kl}$  form a connected path from $i$ to $l$. 
Initially, $V_{s>1}=0$. $V_1$ alone can form only resonant terms in $A_3$  such as $iT_1P_1T_1$, which are of order $\varepsilon^3$.  In the linear early time, $V_3$ is thus of order $\varepsilon^3$. The phase of $\alpha_{3,\pm}$ are ahead of $\alpha_{1,\pm}$ by $\pi/2$ because of the coefficient $i$. $V_1$ and $V_3$ together can form more resonant terms such as $A_5$  of order $\varepsilon^5$ and leads to the same order $V_5$. There are also paths as $1+3-3=1$ or $3-1-1=1$ in $A_1$. These terms are of higher order than $\varepsilon$ and negligible in $A_1$. One by one we can show that, in early time, the dominant terms in $A_s$ is of order $\varepsilon^s$ coming from the shortest path. It leads to a same magnitude of $V_s$. With a similar logic, the initial relative phase between any nearest odd sites is $\pi/2$.  

\begin{figure}[tbp]
    \centering
    \includegraphics[width=.5\textwidth]{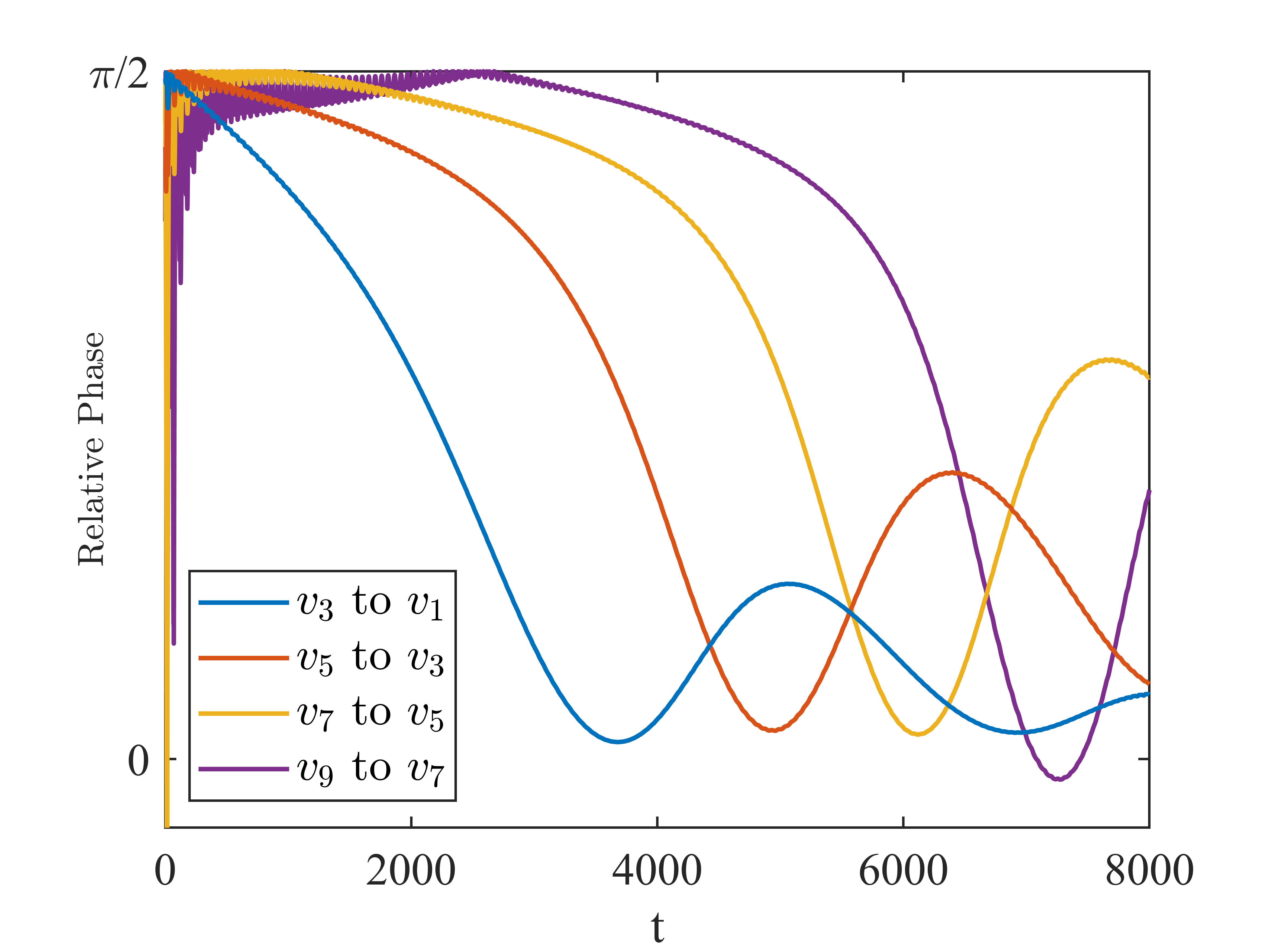}
    \caption{The relative phase between the complex amplitudes of nearest odd sites.}
    \label{f_phase}
\end{figure}

\begin{figure}[tbp]
\begin{subfigure}{.5\textwidth}
  \centering
  \includegraphics[width=.9\linewidth]{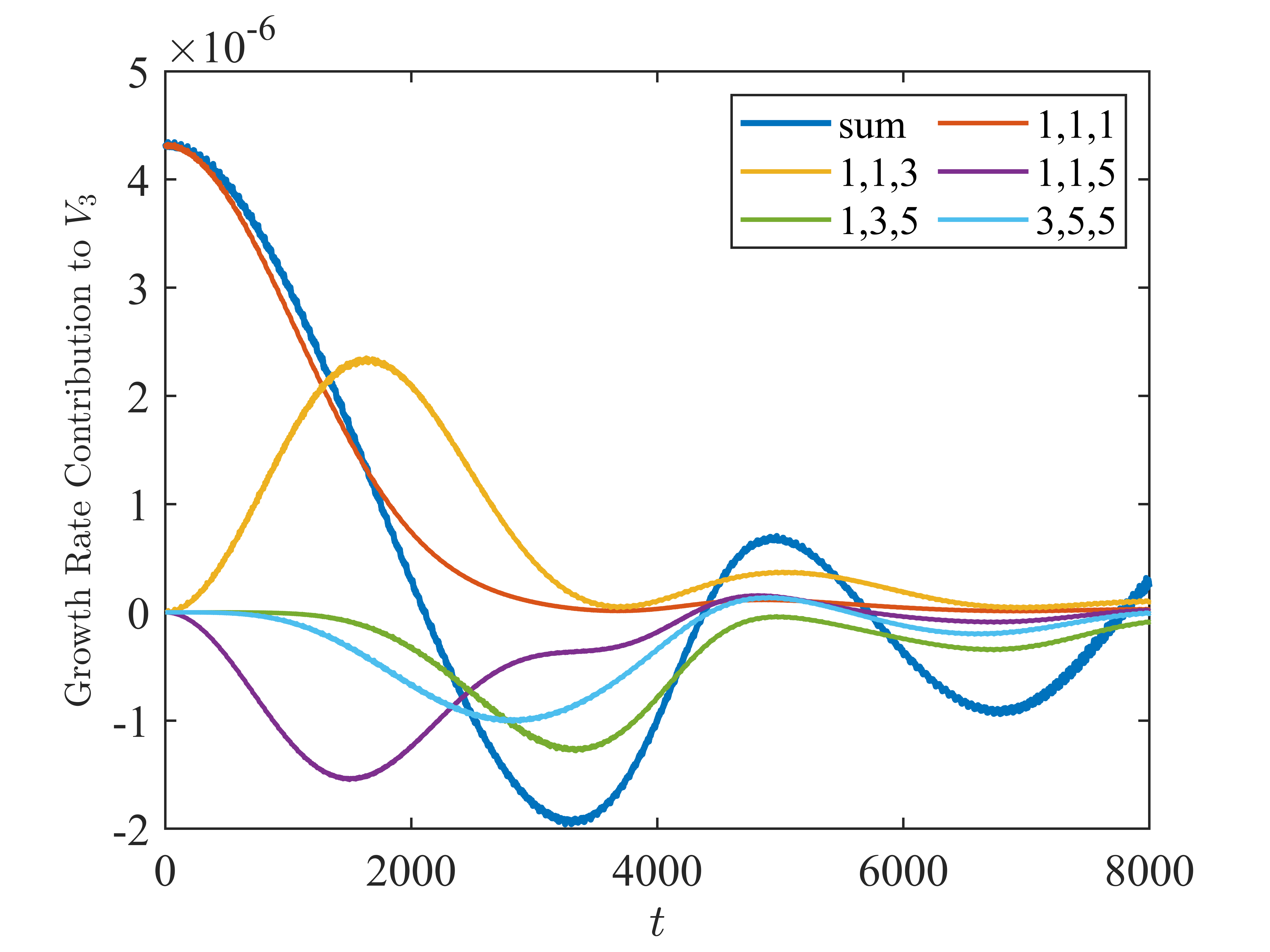}  
  \caption{}
  \label{f_s3}
\end{subfigure}
\begin{subfigure}{.5\textwidth}
  \centering
  \includegraphics[width=.9\linewidth]{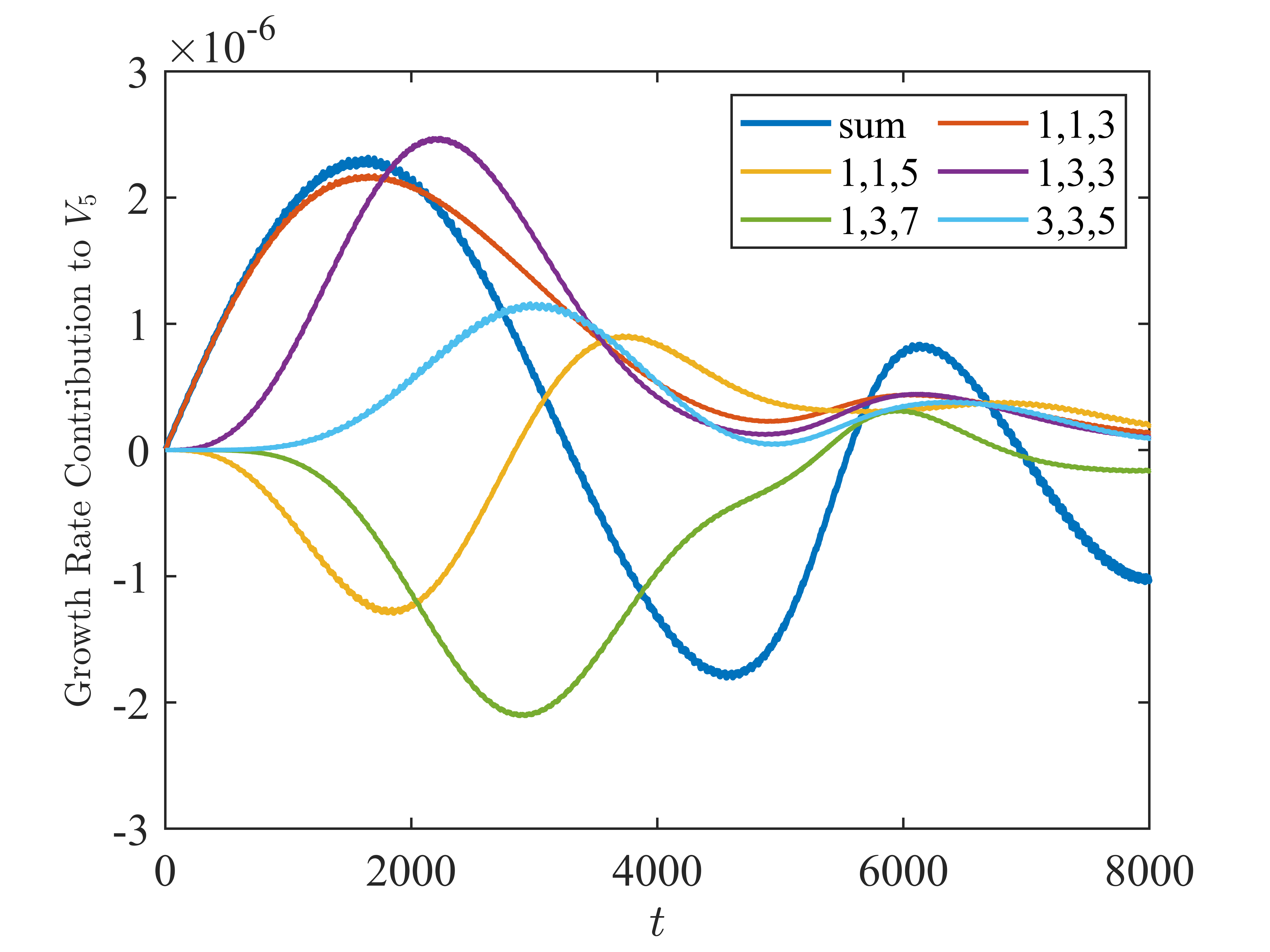}  
  \caption{}
  \label{f_s5}
\end{subfigure}
\caption{The contribution of resonant terms in $A_i$ to the growth rate of $|\alpha_{i,+}|$ in $V_i$. $i=3$ in (a) whereas $i=5$ in (b). The $i,j,k$ lines are the resonant terms within $iV_iV_jV_k$ structures. The 'sum' line is the sum of all resonant terms, many of which are not plotted here.}
\label{f_terms}
\end{figure}

In a later time, $V_3$ grows to order $\varepsilon$ with rate $\varepsilon^3$ under the driving force $A_3$. The dominant terms in $A_5$ thus grow to also $\varepsilon^3$ and arouse $V_5$ to grow at this same rate. The wave propagates in odd collective variables with coordinate speed of order $\varepsilon^2$.\footnote{For even s, $T_aP_bT_c$ or $P_aT_bP_c$ terms in $V_s$ has at least another even $V_{s'}$. The oscillations of $V_0$ around $V_*$ have different frequency. As a result, there are no resonant terms. } 
For a more detailed description of the dynamics, we should consider all possible resonant terms. We plot the relative phase of some nearest odd sites in figure \ref{f_phase}. They begin at $\pi/2$ and descend close to 0 as the corresponding $V_s$ peaks. The former sites of roughly the same phase are composed to driving terms in $V_s$, which are $\pi/2$ ahead of them. As the phase gap approaches 0, the former driving force became a phase term which contribute little to the absolute amplitudes. In the other way around, the latter sites are combined to be damping forces of the previous sites. Before the peak, the driving force dominates. After the synchronization, the driving force dies out and the damping force grows as the latter sites grows. The relative phase between complex amplitudes thus leads to the varying effects of $A_s$. The figure \ref{f_terms} shows the contribution to the time derivative of $V_{s,+}$ from resonant terms in $A_s$. The dark blue line is the sum of all resonant terms, which indeed reflects how $V_s$ changes. For example, the sum in $A_1$ becomes sub-zero at around $t=2000$, where $V_1$ peaks. In figure \ref{f_s3}, initial dominant terms is indeed given by $V_1V_1V_1$ like terms. It approaches to 0 as $V_3$ peaks. While terms involve $V_5$
or higher order terms contribute negatively to the amplitudes. They would also descend close to 0 as $V_5$ peaks. The late time dynamics are dominated by even higher collective variables. Similar things happen in figure \ref{f_s5}, where combinations of $V_1$ and $V_3$ contribute positively but $V_7$ involves negatively. We can see in above, the information wave is physically similar to classical waves. Each odd site is driven by preceding sites and dragged down by succeeding sites.


    \bibliographystyle{JHEP}
    \bibliography{EL}
\end{document}